\documentclass[aps,prl,reprint,groupedaddress]{revtex4-1}
\usepackage{graphicx} 
\usepackage{dcolumn} 
\usepackage{bm}
\usepackage{hyperref}
\usepackage{siunitx}
\usepackage{float}
\usepackage{graphicx} 
\usepackage{epstopdf}
\usepackage{verbatim}
\usepackage{array}
\usepackage{booktabs}
\usepackage{multirow}
\usepackage{amsmath}
\usepackage{amssymb}
\usepackage{setspace}

\begin{document}

\title{Hyperon dynamics and production of multi-strangeness hypernuclei in heavy-ion collisions at 3A GeV }
\author{Hui-Gan Cheng }
\author{Zhao-Qing Feng }
\email{Corresponding author: fengzhq@scut.edu.cn}

\affiliation{School of Physics and Optoelectronics, South China University of Technology, Guangzhou 510640, China}

\date{\today}

\begin{abstract}
Within the microscopic transport, systematic investigation of the many facets of hyperons and hypernuclei up to strangeness $S = -2$ are carried out for $^{197}$Au + $^{197}$Au and $^{40}$Ca + $^{40}$Ca at the incident energy of $3A$ GeV. The spatial, temporal and density distributions of hyperon production, absorption and freeze-out are thoroughly investigated. The rapidity and kinetic energy spectra of $\Xi$ hyperons and double $\Lambda$ hypernuclei are analyzed. It is revealed that the chemical balance of hyperon production is established in baryon-baryon channels while the opposite is found in baryon-meson channels. It turns out that the rapidity spectra of $^{4,5}_{\Lambda\Lambda}$X are single-peak and more than two orders of magnitude lower than that of $^{3}_{\Lambda}$H. Formation of double $\Lambda$ hypernuclei through $\Xi$ hypernuclei as intermediate states is also discussed in kinematics.

\begin{description}
\item[PACS number(s)]
21.80.+a, 25.70.Pq, 25.75.-q             
\end{description}
\end{abstract}

\maketitle

In past decades, great attention has been attracted to the investigation of hypernuclear physics. In astrophysics, the so far blurred theoretical picture and the great experimental uncertainties of multi-strangeness interactions (interactions involving $N$, $Y$, $\Xi$ and $\Omega$ hyperons which are expected to emerge copiously at densities beyond 2-3$\rho_{0}$ \cite{Fo15,Ba14,Pr97}) are the principal obstacles in our way towards a consistent description of the structure of compact stars. The presence of hyperons can either stiffen or soften the equation of state(EOS) of dense nuclear matter depending on the details of hyperon interactions, which poses great uncertainties in the prediction of the maximum mass of neutron stars \cite{Da53,Am08}. In heavy-ion collisions at incident energies below 2A GeV, the staggering enhancement of hyperon production currently not unanimously resolved at hadronic level \cite{Wh09,Ch04,Gr14} may be interpreted as a possible signal for the onset of the formation of a deconfined phase in this energy regime\cite{Ra82,Ra86,Ra91}, which is yet particularly suspect\cite{Su04}. Since hyperons carry nonzero strangeness number and are thus exempt from the Pauli exclusion principle, the addition of hyperons to ordinary nuclei extends the nuclear chart to the strangeness sector\cite{Ba90,Ha06,Bu13,Sc93,Gr96} which is characterized by exotic physical properties, for example, extreme isospin and higer saturation dentisty.

Due to the experimental infeasibility of direct preparation of hyperon targets for their short lifetimes, hypernuclei embedding more than one hyperons are the only crutch available that people have resorted to in probing the hyperon-nucleon and hyperon-hyperon interactions beyond $S = -1 $. There exists a wealth of models including the Effective Field Theory model(EFT) of Bochum/Juelich\cite{Ha05}, chiral-unitary approach of Sasaki, Oset and Vacas\cite{Sa06}, the Nijmegen models\cite{Th99,Th06}, the quark-cluster models\cite{Fu01}, and chiral effective field theory\cite{Ha10}, for example, providing independent descriptions of strangeness interactions which mostly converge at the $S =$ 0 and -1 levels. The disagreement among their predictions at $S = -2$ and beyond calls for increasingly strigent experimental constraints if any serious advance was to be made. However the current status of the experimental research concerning multiply strange hypernucleus production is far from being satisfactory. Up to now, only a handful of events of double lamda hypernuclei have been observed via their double pion decay in reactions induced by hadrons(see \cite{Po05,Ek19} and the references therein). Thus an elaborate setup P\textsc{anda}\cite{Po05} is being planned at FAIR as the next-generation facility for the 'mass production' of double lamda hypernuclei. Meanwhile relativistic heavy ion collisions have proved an abundant source of strangeness\cite{Lo14,Ag11,Ag09,Al08,Ad07,Sc06,An04,Ch03,Al02} and the endeavors to study the production of hypernuclei in this mechanism were started as early as the 70s theoretically by Kerman and Weiss\cite{Ke73}, and in the laboratories at Berkeley\cite{Ni76} and Dubna\cite{Av92}. Today successful measurements of the rapidity and kinetic energy spectra of single-lamda hypernuclei at HypHI\cite{Ra15} have reassuringly confirmed heavy ion collisions to be a promising way for this purpose. Again, when it comes to heavy ion collisions concerning the production of double lamda hypernuclei, however, the current progress is still limited to a very primitive stage, even in terms of advances of theoretical predictions(only studies on production yield exist\cite{Lo11,Bo17}). Thus it would be beneficial, as also attempted in this work within the framework of quantum molecular model, to give an estimation of the kinematic spectra of double lamda hypernucleus production as a valuable theoretical guidance for setups in relevant experiments at facilities like FAIR \cite{Gu06,Ay16,Ge03} in Germany, NICA in Russia \cite{Nica}, J-PARC in Japan \cite{Ta12} and HIAF\cite{Ya13} in China in the near future.

At energies higher than $\sqrt{s_{NN}} > 5$ GeV, the production of strange particles is satisfactorily reproduced by grand canonical analysis of particle yields and by transport simulations\cite{Ch03}, suggesting the establishment of chemical equilibrium. But it is rather inconclusive whether this is also justified at lower energies in that both statistical and microscopic approaches(except for one transport model\cite{Li12} which is neither unquestionable) fail in correctly describing the production of $\Xi$ hyperon\cite{Wh09,Ch04,Gr14} at $1\sim2$A GeV. In face of this, the question naturally arises whether there are some important sources of $\Xi$ hyperon still missing in our present understanding of the reaction mechanism, or it is any lurking features particular to this energy regime that indirectly give rise to this puzzle. Thus a systematic investigation on the various aspects of hyperon production will shed interesting light on the details and characteristics of the process, and help us gain confidence in deciding in the future on the various possible solutions to the problem. Stimulated by this possibility, we carried out a thorough investigation and comparison on the spatial, temporal and density distributions of hyperon production, absorption, and freeze-out within spatial zones of different characteristics(fireball, mixed zones, spectators and free zone) and within reaction channels of different nature(baryon-baryon and baryon-meson), respectively, to ascertain within transport approach such things as the establishment of equilibrium, freeze-out, etc.

In heavy ion collisions at moderate relativistic energies in which it mostly suffices to look no deeper than at the hadronic degrees of freedom, hyperons are mostly produced directly in primary nucleon-nucleon collisions above the thresholds, which are 1.57 GeV or so for $NN \rightarrow BYK$\cite{Ts99} and 3.74 GeV for $NN \rightarrow N\Xi KK$ in the laboratory frame with a cross section of tens to hundreds of $\mu$b depending on the energy available and the species of baryons  produced in the final state.  Here $Y$ denotes ($\Lambda$, $\Sigma$), $B$ ($N$, $\Delta$, $N^{*}$), $\Xi$ ($\Xi^{-}$, $\Xi^{0}$), $\pi$ ($\pi^{-}$, $\pi^{0}$, $\pi^{+}$), $K$ ($K^{0}$, $K^{+}$) and $\bar{K}$ ($\bar{K}^{0}$, $K^{-}$). Also of the same order of magnitude are the cross secions of pion induced channels $\pi N \rightarrow YK$\cite{Sc83} and the antikaon induced channels $\bar{K}N \rightarrow K\Xi$. In the contrary school, the typical cross secions of secondary interactions among strange particles of $S = -1$ themselves to produce $\Xi$ hyperons are enormous, tens of mb\cite{Li12}, say, comparable to that of $NN$ collisions. Another major type of channels whose magnitude of cross sections falls between the two schools, is $N\bar{K} \rightarrow Y\pi$, about several mb above the thresholds\cite{Fl83}. Meanwhile the momentum released in the conversion of the $\Xi N$ system back into the $\Lambda\Lambda$ system is below fermi momentum $270$ MeV/c, thus making possible the simultaneous capture of the two $\Lambda$ if the initial $\Xi$ lies below the fermi momentum. Thereby it would make sense, as will be done in this work, to compare between the contributions to the production of double lamda hypernuclei through the direct capture of two $\Lambda$ and that through the formation of the intermediate(presumably) $\Xi$ hyperfragment.

In previous studies, the production of single lamda hypernuclei has been extensively carried out based on popular transport models like the Dubna Cascade model(DCM)\cite{To83,To90,Bo11,St12}, Giessen BUU model(GiBUU)\cite{Bu12,Ga09,Ga14}, Hadron String Dynamics model(HSD) \cite{Ca99,Su18} and Ultrarelativistic Quantum Molecular Dynamics model(UrQMD)\cite{Ba98,Bo11,St12,Su18} and so on, applying coalescence\cite{Wa88,St12,Bo15,Bo17,Su18} or potential\cite{Bo11,Bo17} criteria, or Minimum Spanning Tree(MST)\cite{Go97,Fe19}, etc. In parallel, the Statistical Multifragmentation Model(SMM)\cite{Bo95} is also modified to include partitions involving hypernuclei\cite{Bo07,Lo11,Bo13}, predicting the formation of hypernuclei beyond the drip lines\cite{Bu13}. As is common, SMM has also been jointly used with transport models to study the decay of the hot fragments in collisions considering strangeness\cite{Su18}. In our current physical picture, light hypernuclei are mainly formed through phase space coalescence in the mid-rapidity fireball region which is the dominant source of strangeness. Hyperons from the fireball region can be slowed down and captured by the spectators if they fall within the spectator rapidity region after multiple scatterings. Since the spectators formed in heavy ion collisons at these energies are endowed with a typical temperature $T\lesssim$ 5-7 MeV\cite{Bo95,Xi97,Sc01}, hypernuclei of various masses and isospins are expected to be produced through spectator fragmentation\cite{Su18}. As is known from existing studies\cite{Bo13,Bo15,Bo17}, there would be no substantial increase in the yields of hyperfragments at incident energies beyond 3-5A GeV. Coincidently, below 2A GeV, detectors are usually very effective for separation of hypernuclei, and, furthermore, at laboratory energies of a few GeV, high-intensity ion beams will become available at facilities like FAIR\cite{Gu06,Ay16,Ge03} at GSI and HIAF\cite{Ya13} in the near future. Experimental production of hypernuclei in the regime of moderate relativistic energy is therefore most economical. Thus in this work, we restrict our attention to only two reaction systems, $^{40}$Ca + $^{40}$Ca and $^{197}$Au + $^{197}$Au at the beam energy of 3A GeV.

The numerical code to be employed in our study is the Lanzhou Quantum Molecular Dynamics (LQMD) Model \cite{Fe11,Fe18}. In this model, all the resonances of masses below 2 GeV are included\cite{Fe18} and all relevant channels involving strangeness production and absorption in this energy regime are also implemented. They are
\begin{gather}
BB\rightarrow BYK, BB \rightarrow BBK\bar{K}, B\pi(\eta) \rightarrow YK, YK \rightarrow B\pi \nonumber \\
B\pi \rightarrow NK\bar{K}, Y\pi \leftrightarrow B\bar{K}, YN \rightarrow \bar{K}NN, \nonumber \\
\bar{K}B \leftrightarrow K\Xi, YY \leftrightarrow N\Xi, \bar{K}Y \leftrightarrow \pi\Xi.
\end{gather}
Moreover, elastic channels are also considered for strangeness, $KB \rightarrow KB$, $YB\rightarrow YB$, $YN\rightarrow YN$, for example and their cross sections are taken from the parameterizations in Ref. \cite{Cu90}. Finally by using the same cross sections as that of the elastic channels, the charge-exchange reactions, $KN \rightarrow KN$ and $YN \rightarrow YN$, are included, for instance, $K^{0}p \leftrightarrow K^{+}n$, $K^{+}n \leftrightarrow K^{0}p$, etc\cite{Fe13}.
In the treatment of particle propagation, the chiral effective Lagrangian and relativistic mean-field are applied in the evaluation of the in-medium potentials of mesons and hyperons. The self-energy of hyperons are assumed to be $n/3$ of that of the nucleons, leading to the dispersion relation,
\begin{gather}
	\omega(\textbf{p}_{i},\rho_{i}) = \sqrt{(m_{H}+\Sigma_{S}^{H})^{2}+\textbf{p}_{i}^{2}} + \Sigma_{V}^{H},
\end{gather}
in which $\Sigma_{S}^{H}=n\Sigma_{S}^{N}/3$ and $\Sigma_{V}^{H} = n\Sigma_{V}^{N}/3$ and $n$ is the number of light quarks in the hyperon. This results in an attractive potential of -32 MeV for $\Lambda$ and -16 MeV for $\Xi$ at $\rho_{0}$. The nuclear scalar part $\Sigma_{S}^{N}$ and vector part $\Sigma_{V}^{N}$ are calculated based on relativistic mean-field model with the NL3 parameters\cite{La97}.

\begin{figure*}
\includegraphics[width=15 cm]{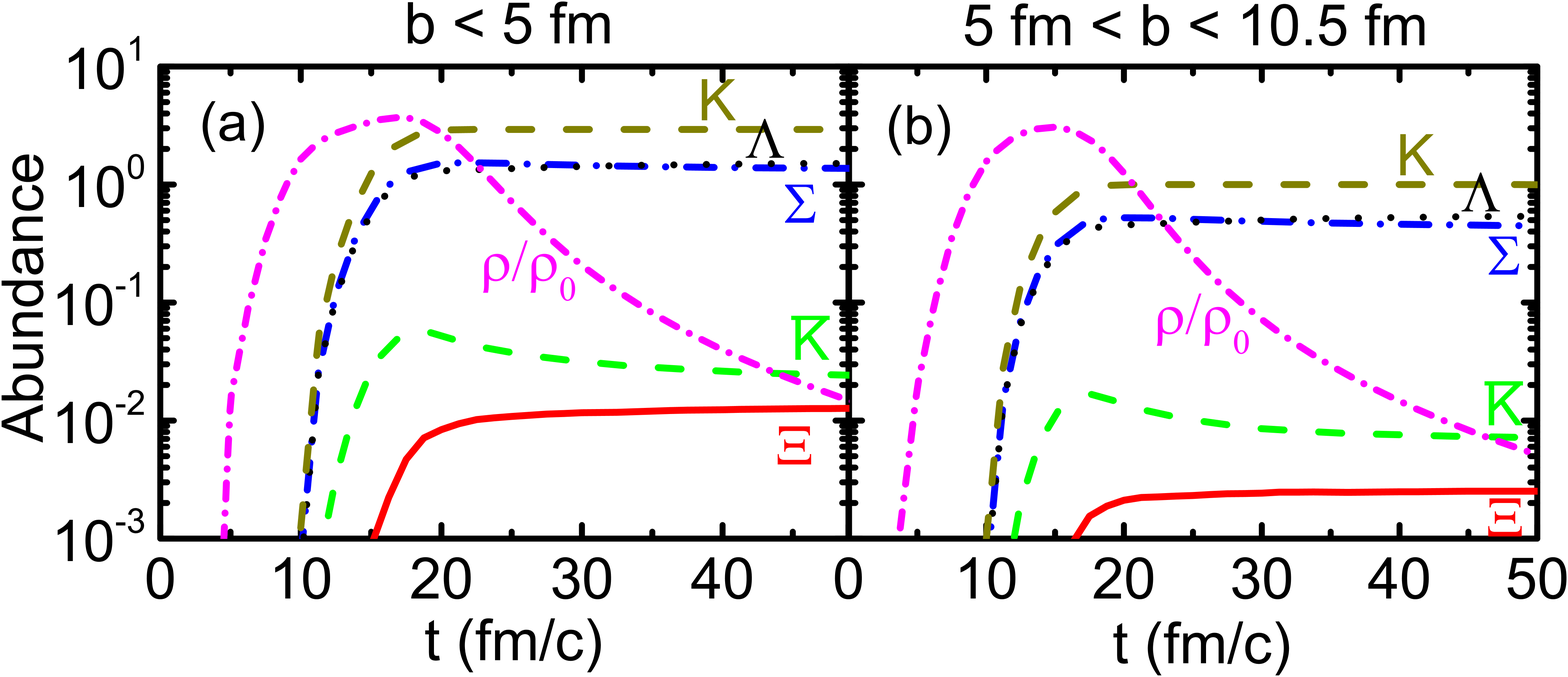}
\caption{Temporal evolution of the abundance of hyperons, kaon and antikaon, and average central density in $^{197}$Au + $^{197}$Au collisions at $3A$ GeV. }
\end{figure*}

 The reaction system in the course of time evolution may be identified as composed of distinct spatial zones, the fireball zone, spectator zones, an overlapping zone bridging in between the two, and finally a free zone, in accordance with the physical picture of Glauber model\cite{Mi07}. We designate the nucleons which live in the longitudinally overlapping areas of the two incident nuclei as fireball particles, whereas the ones outside the overlapping cylinder are spectator particles(see Fig. 3 in Ref. \cite{Mi07}). The identities(fireball particle or spectator particles) of all baryons are held unmodified in the evolution through time, thereby rendering a definition of the four aforementioned zones  possible. For heavy ion collisions at a specific impact parameter, a quantified recognition of these zones requires a close inspection of the time and density information of the interesting particles. Shown in Fig. 1 is the time evolution of central density and particle multiplicities of two different centralities in $^{197}$Au + $^{197}$Au collisions at $3A$ GeV. We observe that in central collisions($b <$ 5 fm as in the left panel), a maximum average baryon density $\rho$ = 4$\rho_{0}$ is reached at 17.5 fm/c (7.5 fm/c after contact between the two incident nuclei) and in peripheral collisions(5 fm < $b$ < 10.5 fm as in the right panel) $\rho$ = 3$\rho_{0}$ at 15 fm/c (5 fm/c after contact). The multiplicity of $\bar{K}$ peaks at 17.5 fm/c in both panels and falls steadily by absorption through $\bar{K} B$ or $\bar{K} Y$ channels. This is accompanied by the steady increase of the multiplicity of $\Xi$ hyperon towards the end of time evolution. From this, one is still unable to conclude whether the chemical equilibrium associated with these particles is established or not since by `equilibrium' we mean that the balance between the production and absorption rates is achieved before the end of time evolution(chemical freeze-out in present study) and all curves will, in principle, reach a plateau after freeze-out of the corresponding particles(freeze-out for the whole system refers to the production and absorption of a particular particle has all ceased). However, it was pointed out in earlier study\cite{Ch04} that the chemical equilibrium associated with $\Xi$ was not established and this is already apparent when we look at the right panel of Fig. 2 where the production and absorption of hyperons through baryon-meson channels are plotted as a distribution against the baryon density at the site of production/absorption. The production and absorption of $\Xi$ through baryon-meson channels are always an oder of magnitude apart from each other, thereby suggesting equilibrium is not established in these channels. It is also shown that the production/absorption of $\Xi$ together with the absorption of $Y$ have broad density distributions which are almost flat between $\rho = 0.4\rho_{0}$ and $\rho = 3.5\rho_{0}$. In comparison, the production of $Y$ through the dominant baryon-baryon channels happens mostly around 2.25$\rho_{0}$ and thus one would assume that they may furnish good probes\cite{Li05} to the properties of nuclear matter within this density region. We will come back to this at the end of our discussion on hyperon dynamics. To conclude this section, we finally remark that our model predicts a ratio $\Xi^{-} / (\Lambda + \Sigma^{0})$ of 0.0035 for central collisions and 0.0020 for peripheral collisions, which are insensitive to centrality but about an order of magnitude lower\cite{Gr14} than what one would anticipate in experiments. Thus the puzzle of hyperon enhancement still persists.

\begin{figure*}
\includegraphics[width=15 cm]{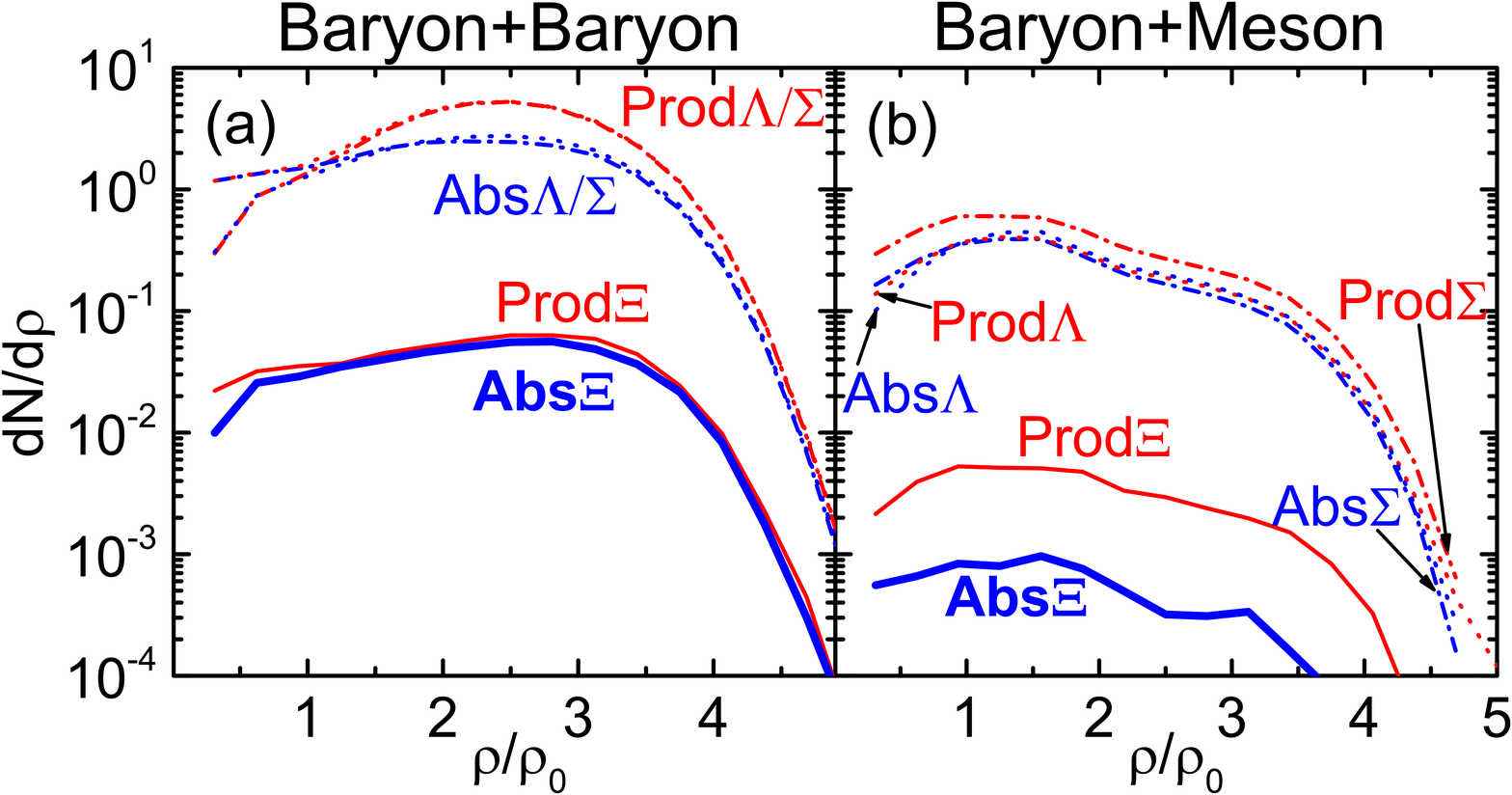}
\caption{Density distribution of production and absorption of hyperons in baryon-baryon channels(left panel) and baryon-meson channels(right panel) in $^{197}$Au + $^{197}$Au collisions at $3A$ GeV and within the centralities of $b =$ 0-10.5 fm. }
\end{figure*}

As was promised, we are to study, in this work, the production of hyperon contributed by distinct spatial zones. To study the chemical freeze-out of hyperons(the very last time a baryon changes its identity), we know from Fig. 1 that it would be safe to terminate the simulation at $t =$ 40 fm/c and from Fig. 2 that it is sufficient to look only at spatial areas of baryon density above 1/8$\rho_{0}$. Thus in the following, we give a quantified definition of these zones whose spatial coverage can be calculated at any moment for a particular reaction system at a fixed impact parameter after an average over many events according to the following designations: Fireball---zone of baryon density $\rho >1/8\rho_{0}$ and $\rho_{F} / \rho > 0.9$ where $\rho_{F}$ is the density of fireball particle as defined in the beginning of the preceding paragraph. Spectator---zones of $\rho > 1/8\rho_{0}$ and $\rho_{S} / \rho > 0.9$ where $\rho_{S}$ is the density of spectator particle as defined. Mixed zone---zones of $\rho > 1/8\rho_{0}$ and not included in the fireball or spectator zones. Free zone---zone of $\rho < 1/8\rho_{0}$. In the light of these designations, the time distribution of freeze-out and production/absorption of different hyperons in different zones are displayed in Fig. 3.

\begin{figure*}
\includegraphics[width=16 cm]{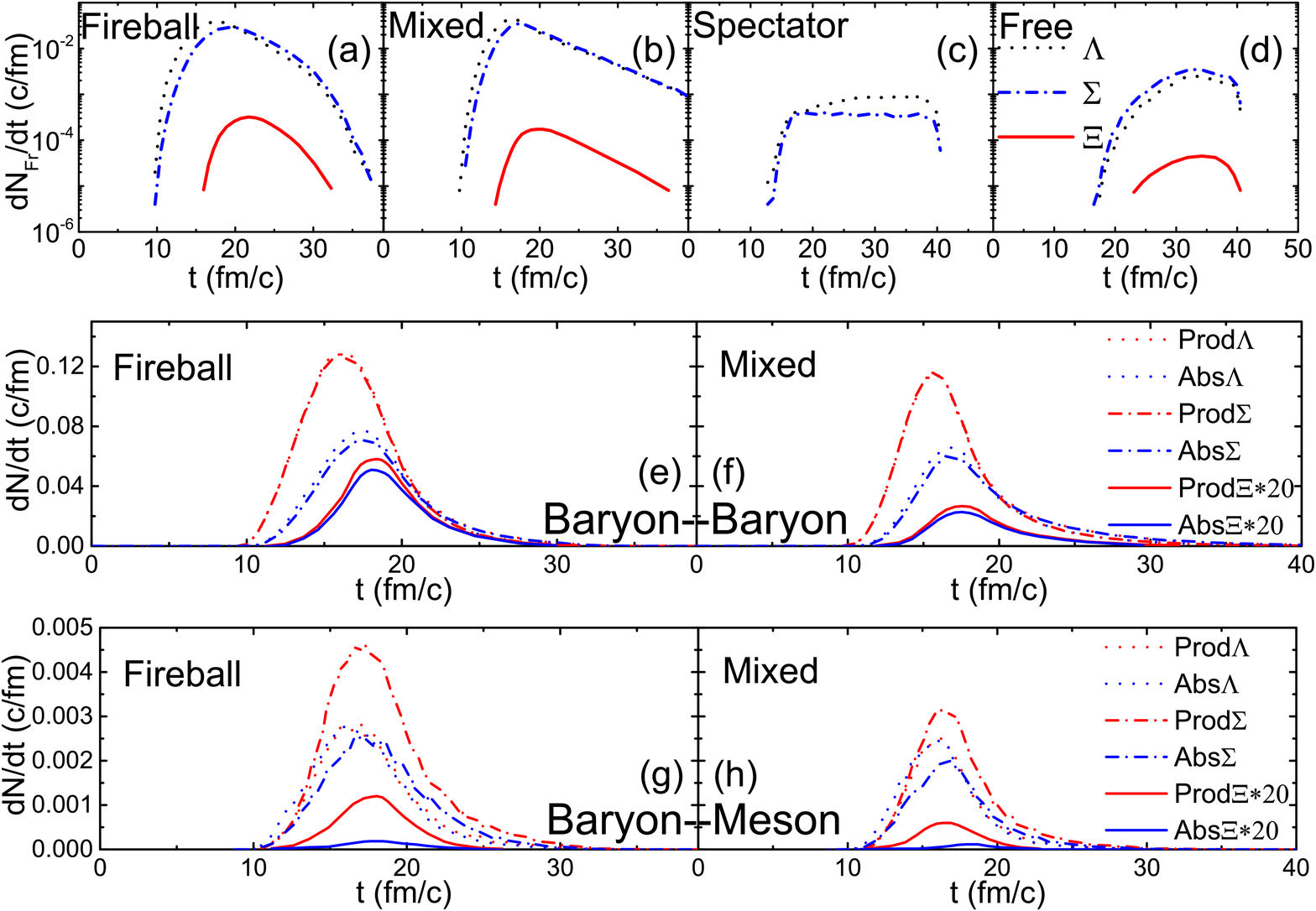}
\caption{Density distribution of production and absorption of hyperons in baryon-baryon channels(left panel) and baryon-meson channels(right panel) in $^{197}$Au + $^{197}$Au collisions at $3A$ GeV and within the centralities of $b =$ 0-10.5 fm. }
\end{figure*}

\begin{figure*}
\includegraphics[width=15 cm]{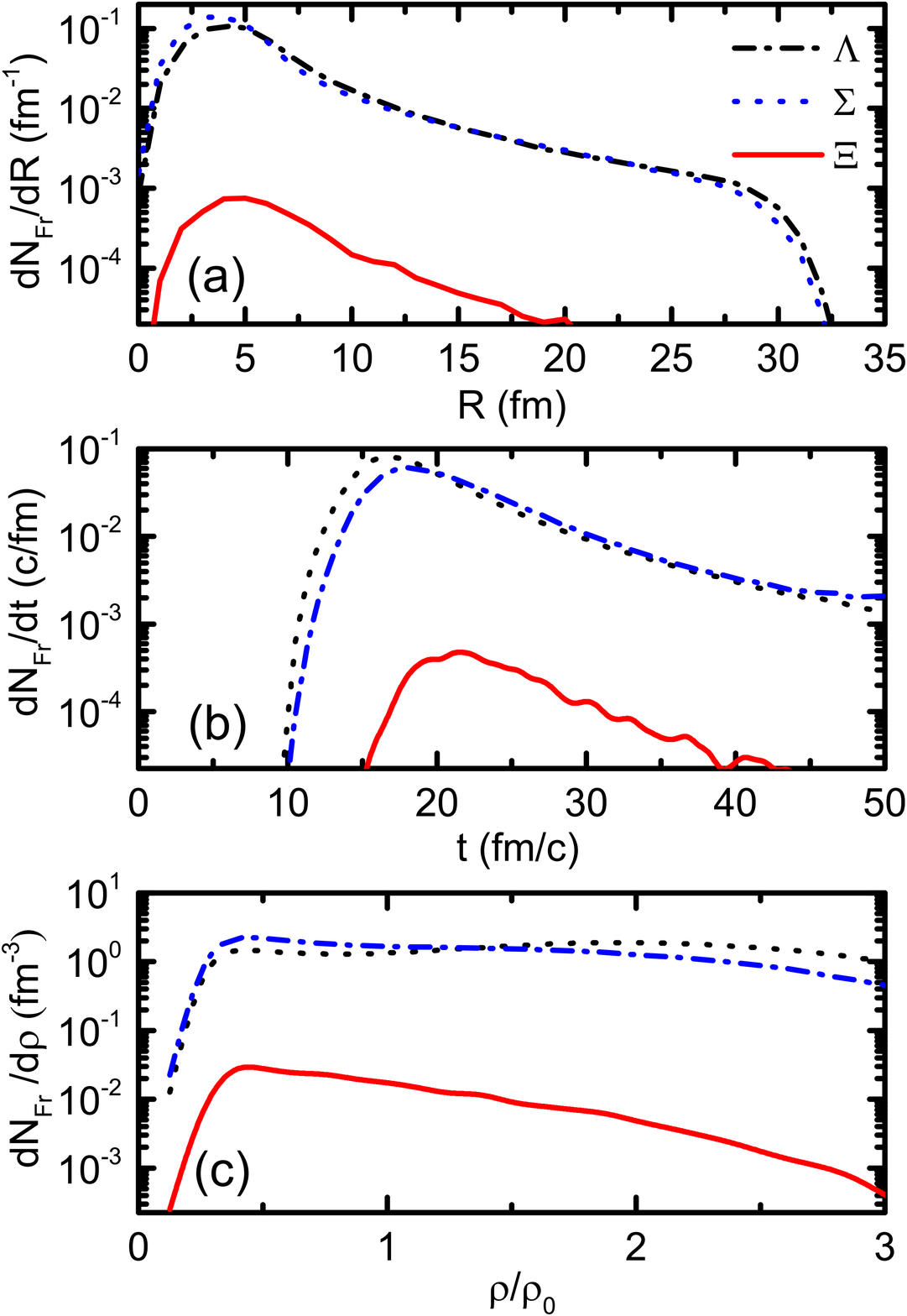}
\caption{Radial, temporal and density distributions of hyperons at freeze-out in $^{197}$Au + $^{197}$Au collisions at $3A$ GeV and within centralities $b =$ 0-10.5 fm. }
\end{figure*}

In baryon-baryon channels, the production and absorption of all hyperons are equilibrated well before the end of time evolution both in the fireball zone and the mixed zones. The corresponding results of the spectator zones and free zones are not presented due to poor statistics. Whereas in the baryon-meson channels, the production of $\Xi$ is always overridingly higher than absorption and this makes it manifest and definite that the equilibrium of $\Xi$ production and absorption is not established, which is comprehensible in that the cross sections of $\bar{K} N \leftrightarrow K\Xi$ channels are small and also the chance of encounter between $K$ and $\Xi$ is remote compared with that between $\bar{K}$ and $N$. However, the final $\Xi$ multiplicity would not be substantially modified since $\Xi$ production through baryon-meson collisions is an order of magnitude lower. In the top panel of Fig. 3, the time distributions of the freeze-out of hyperons in the four zones are presented and, as expected, the patterns of hyperon freeze-out distributions in different zones are distinct from one another. Since the fireball and the mixed zones account for the majority of the production of hyperons of all species, most final hyperons come out free from these zones, the fireball zone characterized by a steep rise-and-fall due to rapid compression and expansion, the mixed zone a lag towards the end of time evolution. In the spectator zone, virtually no $\Xi$ hyperons freeze out owing to the strongly absorptive nature of the spectator matter for $\Xi$, while the freeze-out of $Y$ spans evenly over time, implying the secondary $Y$-producing collisions induced by the shower of particles($Y$, $\bar{K}$, $\pi$, $\Xi$, etc.) from the fireball region which continues robustly since 17 fm/c and ends abruptly at 42 fm/c. Thus there exists a broad window for the spectator to slow the hyperons down and capture them as essential ingredients, and termination of the simulation thus has just come to be sufficient only after 42 fm/c as long as the construction of hot hyperclusters is concerned. As a complement, we now turn to the free zone. Contrary to our preconception, this zone, though dubbed `free zone' for sake of its low baryon density and thus the smaller chance of interactions, is not free at all. Therefore the `open zone' may be a more appropriate name for this zone due to its infinite spatial extension. We observe from the free zone panel that although an order of magnitude lower than that of the fireball and the mixed zone, the freeze-out of hyperons in this zone is appreciable near the end of time evolution. Thus the interactions in the free zone may be regarded as a diluted succession of the violent dynamical fireball phase in the vast open space.

In Fig. 4, the radial, temporal and density distributions of freeze-out of hyperons are presented for all events within centralities $b =$ 0-10.5 fm. We note that the average freeze-out radii locate at about $R =$ 5 fm, a value common to all three species of hyperons, which is in agreement with previous studies\cite{Su04} of incident energy 6A GeV. The region within this radius coincides with the fireball, which, from the transport point of view, justifies the validity of the use of statistical approaches\cite{Wh09} in the description of hyperon production. It is worth noting in the time distribution that the time it takes to reach the peak from zero are the same for all hyperons whereas the curve is displaced by 5 fm/c for $\Xi$ with respect to that of $Y$, indicating that the $\Xi$ hyperons are produced in secondary collisions. Finally we look at the density distribution which is of particular interest. At first glance, it may look a bit puzzling that the freeze-out of $Y$ is quite insensitive to density, and similar thing is true of $\Xi$ within a narrower density range. This may be understood that, though the production rate of $Y$ at high density dominates as mentioned in Fig. 2, the absorption rate is flat in $\rho =$ 0.4-3$\rho_{0}$, in analogy with a constant damping of hyperon abundances in the way of the hyperons traveling through this density interval. Thus only a small fraction of the hyperons produced at high density actually freeze out there. Plus the contribution brought by the production of hyperons in low density regions, a flat density distribution of hyperon freeze-out is thus reasonably expected. The important theoretical consequence of this is that hyperons, despite their copious production at high density, may not serve a promising probe to dense nuclear matter as was demonstrated in our very recent work\cite{Zh21}, due to contamination by their production at lower densities. This is particularly true of the freeze-out of $\Sigma$ and $\Xi$ hyperons which peaks at 0.4$\rho_{0}$ and falls hereafter. If we have to employ hyperons as probes, two helpful remarks are in order. The transverse direction($\theta_{polar}=\ang{90}$) is most favored by the emission of hyperons which have frozen out at high density. By similar analysis based on freeze-out, the high transverse mass tail of the hyperon transverse mass spectra would be a more sensitive probe to the physics at high density compared to the high kinetic energy tail. For the time being, these will be stated without demonstration.

\begin{figure*}
\includegraphics[width=15 cm]{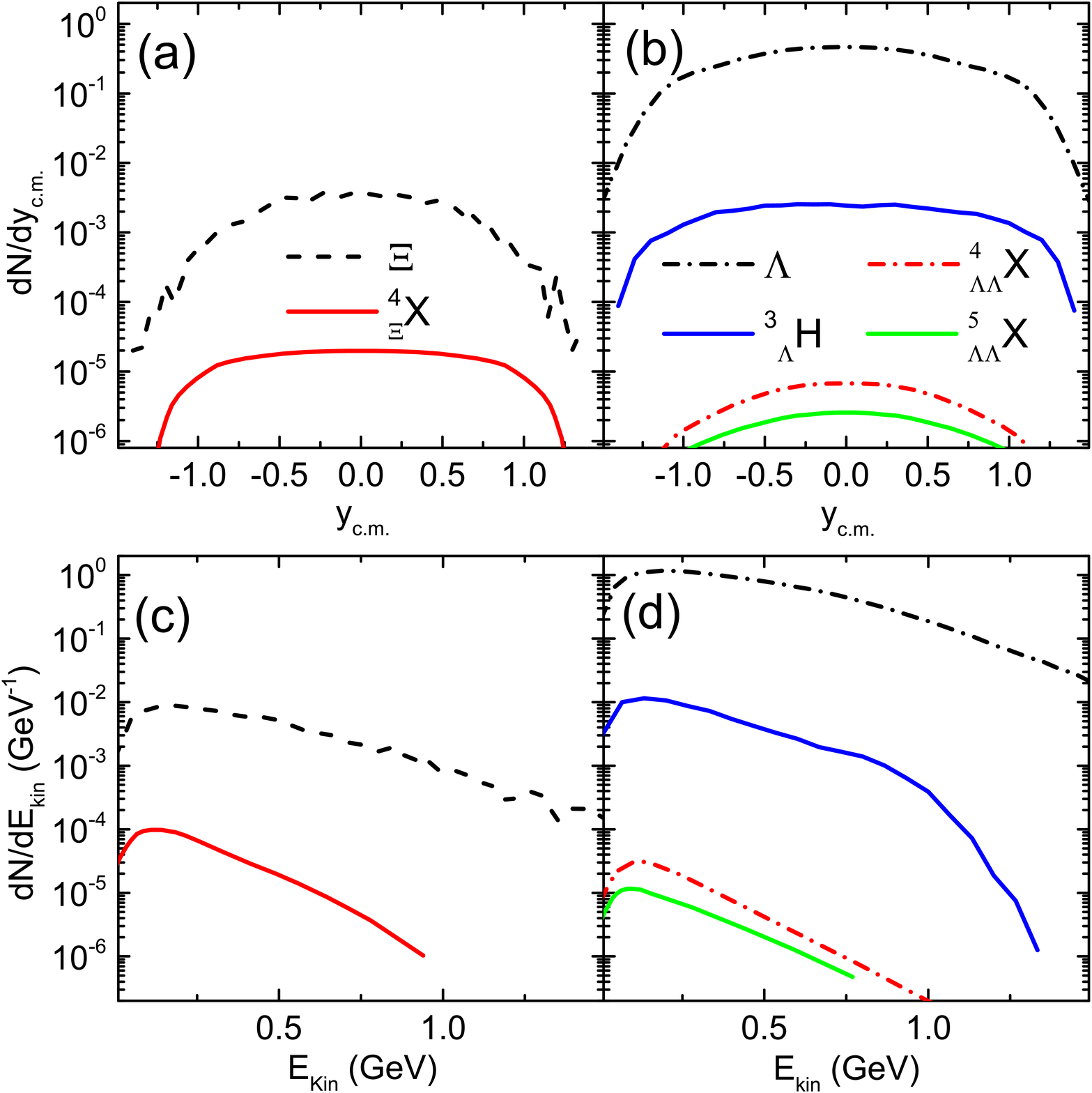}
\caption{Rapidity and kinetic energy spectra of hyperons and hypernuclei up to $S = -2$ in $^{197}$Au + $^{197}$Au collisions at $3A$ GeV within centralities $b =$ 0-10.5 fm, calculated with $r_{0} = 3.5$ fm for $\Lambda$ and $r_{0} = 5.0$ fm for $\Xi$. Here $^{4}_{\Xi}$X includes all single-$\Xi$ fragments of $A = 4$ and $^{4}_{\Lambda\Lambda }$X includes $^{4}_{\Lambda\Lambda}$H and $^{4}_{\Lambda\Lambda}$He. The short-hand $^{5}_{\Lambda\Lambda }$X includes $^{5}_{\Lambda\Lambda}$H, $^{5}_{\Lambda\Lambda}$He and $^{5}_{\Lambda\Lambda}$Li. Note that the in the abscissa of the kinetic energy spectra, we used kinetic energy per baryon for illustrative purposes instead of total cluster kinetic energy. }
\end{figure*}

Having gone through the detailed aspects of hyperon dynamics, it would be a comparatively simple step to go from the analysis of the dynamics of hyperon production to the analysis of the kinematics of hyperon and hypernucleus production up to $S = -2$ as will be carried out in the paragraphs hereon. As is often, the subtle thing about the description of hyperfragment formation in this energy regime, apart from the uncertainties in strangeness production itself, is the complications brought about by the interplay among the creation, transportation and interaction of numerous species of particles. In the violently interacting fireball zone where an appreciable portion of the incident energy is deposited and thus the final-state particles are quasi-free, a coalescence argument based on the overlapping between the cluster Wigner function\cite{Ma97,Fe21} and that of the source may be more complete for light clusters.  Whereas in the spectators left behind after peripheral collisions where the nuclear matter is just moderately excited, say to a temperature of $T\lesssim$ 5-7 MeV, and the forces between hadrons are still prevalent, algorithms like MST, Simulated Annealing Clusterization Algorithm(SACA)\cite{Pu00,Fe19} or more sophisticated approach based on explicit consideration of cluster correlations\cite{Da91,Ku01,On13} should be applied in the construction or description of primary fragments to avoid the overestimation of hypernucleus production in the spectator rapidity region. Thus in this work, for the construction of hypernuclei, the MST is adopted in which the proximity parameters $r_{0} = 3.5$ fm in coordinate space and $p_{0} = 200$ MeV/c in momentum space are employed respectively for nucleon-nucleon coalescence. For coalescence involving hyperons, we take $r_{0} = 3.5$ fm and also another choice $r_{0} = 5.0$ fm is ventured.

In Fig. 5, the rapidity and kinetic energy spectra of $\Xi$ and $\Lambda$ hyperons, $^{4}_{\Xi}$X, and double lamda hypernuclei of $A = 4$ and $A = 5$ calculated with $r_{0} = 3.5$ fm for $\Lambda$ and $r_{0} = 5.0$ fm for $\Xi$, are plotted and compared for $^{197}$Au + $^{197}$Au at $3A$ GeV within $b = 0\sim10.5$ fm. In this, we call for attention to the units we adopted in the abscissa of kinetic energy spectra where we have used kinetic energy per baryon for illustrative purposes in stead of total kinetic energy. Anyway, the total yields are the areas below the curves in the plots. We first note that the rapidity spectra of $\Lambda$ hyperons(without adding the contribution from $\Sigma_{0}$ decay) is in reasonable agreement with the recent result calculated within PHQMD\cite{Ai21} and the rapidity spectra of $\Xi$ is steeper than that of the former and two orders of magnitude lower. The same is true of their kinetic energy spectra. This comes as no surprise since $\Xi$ is produced in later collisions along the cascade of energy deposition. Particular emphasis should be put on the kinematics of double lamda hypernuclei $^{4}_{\Lambda\Lambda }$X and $^{5}_{\Lambda\Lambda }$X. The former is short for the sum of the contributions of $^{4}_{\Lambda\Lambda}$H and $^{4}_{\Lambda\Lambda}$He, while the latter is short for the sum contributed by $^{5}_{\Lambda\Lambda}$H, $^{5}_{\Lambda\Lambda}$He and $^{5}_{\Lambda\Lambda}$Li altogether. For comparison, also displayed is the results of $^{4}_{\Xi }$X short for all primary clusters of $A=4$ and with a $\Xi$ embedded which may act as possible intermediate states before the finally observed double lamda hypernuclei. Note that the rapidity spectrum of $^{4}_{\Xi }$X rates about the similar order of magnitude as that of $^{4}_{\Lambda\Lambda }$X.  We learn from these plots that the rapidity spectra of all the double lamda hypernuclei up to $A = 4$ peak around the central rapidity region and contains a long tailing-off down into the spectator rapidity regions($y=\pm1$), within the same order of magnitude as in the central ones. Apparently, this is a strong indication that although coalescence in the mid-rapidity fireball region is the dominant formation mechanism of $^{4}_{\Lambda\Lambda }$X and $^{5}_{\Lambda\Lambda }$X, the contribution from the spectators is equally important. This is in striking contrast with the double peaks that are located around the spectator rapidity regions for small or medium-sized reaction systems, $^{40}$Ca + $^{40}$ca at $3A$ GeV for example. We have in fact confirmed the double-peak structure for $^{40}$Ca + $^{40}$Ca by carrying out explicit calculations but, unfortunately,  the statistics is too low to enable a smooth plot to be drawn, as will be appreciated when we come to the comparison between the cross sections in table \ref{cross_sections}. Of course, the double-peak distribution is already observed at HypHI\cite{Ra15}, which can only be accounted for by assuming coalescence around the spectator as dominant source of hypernuclei\cite{Su18,Fe19,Fe20}. As was concluded in Ref.\cite{Bo17}, the production of hypernuclei decreases by two orders of magnitude or more consecutively with the addition of more and more $\Lambda$ hyperons. Due to low production, it would be more sensible, in experimental practice, to choose heavy reaction systems if we temporarily disregard other conditions such as detector efficiency and acceptance, associated with the instruments. Finally we note in passing that the above results are calculated with $r_{0} = 3.5$ fm for $\Lambda$ and $r_{0} = 5.0$ fm for $\Xi$ in the MST. If $r_{0} = 5.0$ fm other than $3.5$ fm is employed for coalescence involving $\Lambda$, the production of double lamda hypernuclei will be enhanced by $2\sim3$ times varying according to their mass number.

To complete our discussion, we now turn to table \ref{cross_sections} for the total yields of light hypernuclei produced in the two reaction systems. These are calculated with $r_{0} = 3.5$ fm for $\Lambda$. In the first place, we note that the cross sections given in Ref. \cite{Bo17} for double lamda hypernuclei in $^{208}$Pb + $^{208}$Pb are the total yields of double hypernuclei, light and heavy, and amount onto the level of 1 mb. Confronting our results of $^{197}$Au + $^{197}$Au with this, it seems our model is underestimating, but in fact, the contribution of heavier double hypernuclei which we have temporarily omitted from the table is huge(bigger clusters are easy to form in the spectators and easier to have two or more hyperons captured in them). We would retrieve a similar result as that of Ref. \cite{Bo17} when summation is performed over the mass number. Finally, one observes in table \ref{cross_sections} that the yields of $^{40}$Ca + $^{40}$Ca are two orders of magnitude lower than that of $^{197}$Au + $^{197}$Au, which is also obvious in Ref. \cite{Bo17} in terms of total $^{A}_{\Lambda\Lambda}$X yields.
\begin{table}[h]
	\vspace{20pt}
	\centering
	\caption{Comparison between cross sections of double lamda hypernuclei calculated with $r_{0} = 3.5$ fm for $\Lambda$ in $^{197}$Au + $^{197}$Au and $^{40}$Ca + $^{40}$Ca collisions at $3A$ GeV}
	\begin{spacing}{1.4}
	\setlength{\tabcolsep}{1.0em}
	\begin{tabular}{lcc}		
		\hline
		\specialrule{0em}{1pt}{1pt}
		Hypernuclei & \multicolumn{2}{c}{Cross sections (mb)} \\ \cline{2-3}
		\hline
		\specialrule{0em}{1pt}{1pt}		
		& $^{197}$Au + $^{197}$Au & $^{40}$Ca + $^{40}$Ca \\
		\hline
		\specialrule{0em}{1pt}{1pt}
		$^{4}_{\Lambda\Lambda}$H	& $2.6\times10^{-2}$ & $1.0\times10^{-4}$ \\ 
		$^{4}_{\Lambda\Lambda}$He	& $1.0\times10^{-2}$ & $\sim10^{-5}$ \\ 
		$^{5}_{\Lambda\Lambda}$H	& $5.9\times10^{-3}$ & $\sim10^{-5}$ \\ 
		$^{5}_{\Lambda\Lambda}$He	& $5.1\times10^{-3}$ & $\sim10^{-5}$ \\ 
		$^{5}_{\Lambda\Lambda}$Li	& $1.4\times10^{-3}$ & $\sim10^{-6}$ \\ 
		$^{6}_{\Lambda\Lambda}$He	& $2.2\times10^{-3}$ & $\sim10^{-6}$ \\ 
		$^{7}_{\Lambda\Lambda}$He	& $6.8\times10^{-4}$ & $\lesssim10^{-6}$ \\ 
		\hline
		
		\label{cross_sections}
	\end{tabular}
	\end{spacing}
\end{table}

In summary, within the framework of LQMD transport model, the hyperon dynamics and hypernuclei production up to $S = -2$ are carried out. It is found that the chemical balance in the bayron-baryon channels involving $\Xi$ hyperon production is established while the opposite was found in the baryon-meson channels and thus the typical reaction timescale is too short in comparison with that needed for equilibration in the baryon-meson channels. More importantly, we have made the improvement by basing our study of the physical processes of hyperon production and reabsorption in the fireball, mixed zones, spectators and free zone on a quantified identification prescription which is not only academically interesting per se but also helps clarify the roles played by the different regions and their interplay. The fireball is characterized by a steep rise-and-fall in the time distribution of freeze-out due to rapid compression and expansion, while the mixed zone exhibits a milder behavior. The contributions from these two zones in hyperon production(freeze-out) are comparable with each other and dominant. On the other hand, the spectators produce hyperons mainly by continuous bombardment of particle shower from the central region which has a duration of about 20 fm/c. Meanwhile there is practically no net production(freeze-out) of $\Xi$ in the spectator zones and the free zone accounts for the majority of the late-stage production of hyperons. In additional discussions, we have confirmed and illustrated, by the density distributions of freeze-out, that indiscriminate use of hyperons as probes to the properties of dense nuclear matter is not advisable and their use as probes would be competent only if certain efficient filtering scheme exists to single out hyperons from high density which we recommend transverse mass spectra around $\theta_{polar} = \ang{90}$. When it comes to the kinematics of hyperons and hypernuclei up to $S = -2$, it is revealed that the rapidity spectrum of $\Xi$ is steep and two orders of magnitude lower than that of $\Lambda$. The rapidity spectra of double lamda hypernuclei $^{4,5}_{\Lambda\Lambda }X$ is more than four orders of magnitude lower than that of $\Lambda$ and the same is true of $^{4}_{\Xi} X$ which may exist as intermediate states before  $^{4,5}_{\Lambda\Lambda }X$. The rapidity spectra of $^{4}_{\Xi} X$ and  $^{4,5}_{\Lambda\Lambda }X$ are all single-peak, flat and centered around mid-rapidity in $^{197}$Au + $^{197}$Au at $3A$ GeV. This is in stark contrast with that of $^{40}$Ca + $^{40}$Ca in which spectator coalescence is the prevailing mechanism of hypernucleus formation. Among other things it was found that the yields of double lamda hypernuclei have a moderate dependence on the coalescence parameter $r_{0}$ for $\Lambda$ and can vary $2\sim3$ times when $r_{0}$ goes from $3.5$ fm to $5.0$ fm. Finally the yields of light double lamda hypernuclei are thoroughly analyzed.

\section{Acknowledgements}
This work was supported by the National Natural Science Foundation of China (12175072, 11722546) and the Talent Program of South China University of Technology.

\end{document}